\documentclass[12pt]{article}
 \usepackage[utf8x]{inputenc}
 \usepackage{graphicx,fullpage,multicol,amssymb,amsbsy}
 \usepackage{setspace}
 \usepackage{epsfig}
 \usepackage{amsmath, amsthm, amssymb, bm,mathtools}
 \usepackage{pgf,pgfarrows,pgfnodes,pgfautomata,pgfheaps}
 \usepackage{color}
 \usepackage{multirow}
 
\usepackage{graphicx}
\usepackage{wrapfig}
\usepackage[round]{natbib} 
\usepackage{verbatim} 

\def\independenT#1#2{\mathrel{\setbox0\hbox{$#1#2$}%
\copy0\kern-\wd0\mkern4mu\box0}}

\title{Bayes-raking: Bayesian Finite Population Inference with Known Margins\footnote{This work was supported by the National Science Foundation grants [SES1760133, SES1534400]}\footnote{Abstract Word Count: 160; Main Text Word Count: 6500}}
\author{Yajuan Si\footnote{Research Assistant Professor, Survey Research Center, Institute for Social Research, University of Michigan; Corresponding author email: yajuan@umich.edu} and Peigen Zhou\footnote{Ph.D. student, Department of Statistics, University of Wisconsin-Madison}}
\begin{document}
\maketitle

\begin{abstract}
Raking is widely used for categorical data modeling and calibration in survey practice but faced with methodological and computational challenges. We develop a Bayesian paradigm for raking by incorporating the marginal constraints as a prior distribution via two main strategies: 1) constructing solution subspaces via basis functions or the projection matrix and 2) modeling soft constraints. The proposed Bayes-raking estimation integrates the models for the margins, the sample selection and response mechanism, and the outcome as a systematic framework to propagate all sources of uncertainty. Computation is done via Stan, and codes are ready for public use. Simulation studies show that Bayes-raking can perform as well as raking with large samples and outperform in terms of validity and efficiency gains, especially with a sparse contingency table or dependent raking factors. We apply the new method to the Longitudinal Study of Wellbeing study and demonstrate that model-based approaches significantly improve inferential reliability and substantive findings as a unified survey inference framework.

\end{abstract}
%%%%%%%%%%%%%%%%%%%%%%%%%%%%
\section{Introduction}
\label{motivation}

Modern survey inference relies on calibration to external control information (e.g., census records) to achieve representativeness of the target population. Calibration weighting, either randomization-based or prediction model-based, has been a popular method to adjust for selection bias due to unit nonresponse and/or coverage errors~\citep{modelass-sarndal92,gem:Folsom,Kott-SM06,calibration:kott09}. Multilevel regression and poststratification (MRP, \cite{park:gelman:bafumi-04,Ghitza:gelman-13,BNFP:SI15,prior-si2018}) connects weighting and prediction via calibration to the population distribution of the auxiliary variables. The rapidly emerging nonprobability-based surveys demand rigorous adjustments for selection bias under potentially informative sampling schemes. The generality of nonprobability survey inference relies on calibration with benchmark information from census or a reference probability survey. The key to calibration is solving the calibration equations that accounts for the auxiliary control information. However, in practice, only marginal distributions of the auxiliary variables or their approximations are known, restricting the quality of the control information and potentially the validity of survey inferences, across design-based, model-based or model-assisted approaches~\citep{model-assist-review-SS17}.

%the use of raking
Raking~\citep{deming1940least} often acts as the final step in survey weighting adjustment to match the unbalanced marginal distributions of variables that affect inclusion (selection and response) from the sample to the benchmark control information, and generally assists categorical data modeling, such as loglinear models, with known margins. Recently raking is widely used to incorporate auxiliary variables that are available to calibrate statistical inference, for example, to analyze missing data~\citep{Little-MDBook02} and evaluate causal relationships~\citep{Schouten18}. Borrowing benchmark or auxiliary data from external sources improves the efficiency and reliability of survey inference, in particular when the raking factors are highly correlated with the survey outcome. While utilizing the iterative proportional fitting (IPF) algorithm for fast computation, raking is faced with methodological and computational challenges due to the wide use and contemporary survey practice. Motivated by survey operation, we propose new strategies to improve raking in terms of variable selection, convergence guarantee and inferential efficiency.

%Problems of IPF
Practical raking often adjusts sociodemographics, the marginal distributions of which are easily accessible. Nevertheless, to make the design ignorability assumption plausible at least conditionally, all variables and their potential high-order interactions that affect the sampling inclusion should be adjusted~\citep{rubin83-pi,little83-pi,gelman14bda}. Raking can compensate the afore-implemented weighting steps, such as inverse inclusion propensity score weighting and nonresponse adjustment, which usually causes high variability with extreme weight values. Weight trimming reduces variance but creates an imbalance of the marginal distributions, potentially increasing the bias of the population inference. The discrepancy relies on raking to adjust a large number of raking variables that affect selection, coverage, and response propensities, and the variance estimation after raking has to account for the design features and weights~\citep{Kott-SM06}, which is available through survey data analysis software such as SUDAAN and the {\em survey} library in R~\citep{sudaan, lumley2016package}. 

The contingency table constructed by the cross-tabulation of numerous raking variables is often sparse or includes empty cells, which can be identified by comparing with the population information if available, and IPF will suffer from a computational burden and fail to converge to the population margins. When all the sample cell counts are nonzero, \cite{ireland1968contingency} prove that IPF can converge with the discussed rate. However, when empty cells occur, convergence can only be achieved under certain conditions and sampling schemes that are not always satisfied~\citep{goodman1968analysis,bishop1969incomplete,fienberg1970iterative}. \cite{raking:brick03} and \cite{battaglia2013practical} have reviewed the problems of raking in survey practice. They find that inconsistencies in the control totals, correlated raking factors, sparse tables, and measurement bias can cause severe issues in raking such as convergence difficulty and high variability. From the theoretical perspective, the inference after raking conditional on empty cells treats the corresponding population cells as empty is not consistent with the design or response mechanism and thus results in estimation bias, in particular for domain inference. Raking adjusts a table to given marginal totals and preserves the interaction structure---defined by the cross-product ratios---of the observed sample table. The retained associations are based on the sample and then subject to sampling variability. With small sample sizes, it is possible that the sample associations deviate from the population associations, resulting in high variability and inconsistent inference. Unconditional analysis that accounts for the variation of the sample cell sizes due to the randomness of the sample inclusion should be performed to yield valid inferences~\citep{Heeringa15}. 

Moreover, correlated raking variables can exacerbate the inconsistency phenomenon. When the main effects are determined by the dependent marginal totals, the observed sample correlation may not be appropriate to reflect the true correlation, and IPF will never converge. \cite{raking:brick03} give an example of age and grade in a two-dimension contingency table, and the correlation causes substantial variation in the raking adjustment factors for the cells even though none of the margins differ much between the sample and the control. With small samples, the raking adjustment factors are possibly below 1 for both age and grade according to the marginal totals. However, the observed sample cell count may be larger than that for the population cell, leading to the failure of IPF. Meanwhile, raking often adjusts high-order interaction terms that are treated as independent ignoring the hierarchy of the high-order interaction terms, which will cause estimation bias and inconsistency.

%Bayesian IPF
We would like to develop a Bayesian framework for raking that settles IPF convergence issues, stabilizes domain inferences, and integrates all sources of uncertainty to ensure inferential validity and efficiency while accounting for complex design and response features and incorporating marginal constraints. \cite{Greenland07,Kunihama13,reiter:ba16} incorporate marginal prior information by appending synthetic observations or pseudo-records to the original data, and the degree of prior uncertainty is controlled by the size of the augmented sample. However, such data augmentation is computationally demanding and not operationally feasible with a large number of raking factors. 

As an innovation, we transfer the marginal constraints as prior distributions and induce to the model of the inclusion indicator. Current methods omit cells with zero sample sizes and can be stuck by boundary estimates with a sparse table. It is nontrivial to adjust the degree of freedom due to empty cells and correct for the estimation bias~\citep{bishop75}. Moreover, the existing approaches cannot handle the issues with raking such as inconsistencies in the control totals or correlated raking factors. We propose to treat the known margins as realizations drawn from the true models and account for the correlation between raking factors based on the hierarchical structure. 

\cite{gelman95bda} proposes Bayesian IPF, a Bayesian analog of IPF, and \cite{schafer:97,Little-MDBook02} present detailed discussions about the properties and examples. Focusing on the estimation of loglinear models with a cross-tabulated contingency table, Bayesian IPF induces a Dirichlet prior distribution to the cell probabilities and generates draws of loglinear model parameters from the fully conditional Dirichlet distributions. \cite{schafer:97} points out that the Dirichlet prior is naive because the cell probabilities are treated as unordered and the cross-specified structure of the contingency table is ignored. However, the cross-classified structure is fundamental to the quality of loglinear models. Alternative Dirichlet prior distributions have been proposed to reasonably incorporate prior distributions utilizing the structure of loglinear model~\citep{Good1967,bishop75}. Normal prior distributions are discussed by \cite{Laird78,Speed88}, although conceptually attractive, but not widely used due to computational difficulty.~\cite{lazar2008} develop a noninformative Bayesian approach to finite population sampling using auxiliary variables, and the proposed prior distribution accounts for the cross-classified structure. Our new methods improve the computation and develop a general subspace format that covers the estimator in~\cite{lazar2008} as a special case, which cannot handle a large number of free parameters.

The proposed Bayesian procedure to incorporate known margins into modeling, that is, Bayes-raking, can solve the problems of the classical raking discussed above. Moreover, Bayes-raking will be embedded and is crucial to our goal of building a Bayesian, design-adjusted and model-assisted framework via MRP that can yield robust survey inferences based on complex or nontraditional design and nonresponse mechanisms including nonprobability samples. MRP starts from the cell estimates of the contingency table constructed by the discrete variables that affect sample inclusion and poststratifies with control information to yield the domain or overall population inference with weights proportional to population cell sizes. Inside each cell, individuals are assumed to be included with equal probability. The multilevel models used for cell estimation stabilize small area estimation~\citep{rao15}, and the poststratification cell structure is robust against model misspecification~\citep{little91, little93, gelman:little-97,gelmancarlin01}. Currently, MRP assumes the population poststratification cell sizes---the joint distribution of calibration variables---are known, rather, only marginal distributions of which are usually given in practice. 

Suppose $\theta$ is the estimand of the survey outcome $Y$, such as the population mean or domain mean. With Bayes-raking, we propose the MRP estimator in the general form,
\begin{align}
\label{model-based}
    \tilde{\theta}^{\textrm{PS}}=\frac{\sum_{j\in D}\hat{N}_j\tilde{\theta_j}}{\sum_{j\in D}\hat{N}_j},
\end{align}
where $\tilde{\theta}_j$ is the expectation estimate, $\hat{N}_j$ is the population cell size estimate in cell $j$, and $D$ is the domain of interest. We will use Stan for the fully Bayesian inference that generates posterior samples of $\hat{N}_j$'s, $\tilde{\theta}_j$'s and $\tilde{\theta}^{\textrm{PS}}$. 

As open source and user-friendly software, Stan improves the posterior computation with nonconjugacy and advocates the model-based survey inference~\citep{stan-manual:2013,stan-software:2013}. We evaluate the Bayesian procedure with frequentist randomness properties as calibrated Bayes~\citep{CalibratedBayes:Little11}. Bayes-raking solves the IPF computational problems, propagates variation and achieves inferential stability with sparse contingency tables under the hierarchical structure and informative prior distributions. We find that Bayes-raking performs at least equivalently to classical raking with large sample size and outperforms when IPF suffers from computation problems. 

The paper is organized as below. Section~\ref{method} starts from a two-variable illustration and describes the details of the proposed Bayes-raking estimation with three main components: the margin model, the inclusion model, and the outcome model. We compare Bayes-raking with raking via simulation studies in Section~\ref{simulation} and apply to the LSW survey in Section~\ref{application}. Section~\ref{discussion} concludes the contributions and further extensions.

%%%%%%%%%%%%%%%%%%%%%%%%%%%%
\section{Bayes-raking}
\label{method}

As an illustration, we consider a two-way contingency table constructed by two categorical variables $A$ and $B$ in the survey data of size $n$, with the marginal distributions of $A$ and $B$ available from a census of size $N$. Let $N_{ab}$ be the population cell count with $(A=a, B=b)$, $N_{a+}$ and $N_{+ b}$ be the known margins, and $n_{ab}$ be the corresponding sample cell count, for $a=1,\dots, d_A$ and $b=1,\dots, d_B$. Here $d_A$ and $d_B$ denote the total number of levels for variables $A$ and $B$.
The sample cell counts $n_{ab}$ are adjusted to estimate population cell count $\hat{N}_{ab}$ to the known margins $N_{a+}$ and $N_{+ b}$, such that 
\begin{align}
\label{margins-2}
    &\sum_b \hat{N}_{ab}=N_{a+}, \mbox{ } a=1,\dots, d_A;
    &\sum_a \hat{N}_{ab}=N_{+ b}, \mbox{ } b=1,\dots, d_B.
\end{align}

\cite{itf:stephan42} shows that the raking estimates of the population cell counts $\hat{N}_{ab}^{Rake}$'s have the form,
\begin{align}
\label{rake-model}
\log(\frac{\hat{N}_{ab}^{Rake}/N}{n_{ab}/n})=\hat{\mu} + \hat{\alpha}_a + \hat{\beta}_b,
\end{align}
with suitable choices of $\hat{\mu}$, $\hat{\alpha}_a$ and $\hat{\beta}_b$. The raking estimates minimize the Kullback-Leibler (KL) divergence $\sum_a\sum_b\hat{N}_{ab}^{Rake}/N\log[(\hat{N}_{ab}^{Rake}/N)/(n_{ab}/n)]$ subject to the marginal constraints in~\eqref{margins-2}~\citep{ireland1968contingency}.

Moving from the discrimination information, \cite{littlewu91} consider four different distance functions of $(\hat{N}_{ab}/N, n_{ab}/n)$ and develop the equivalent maximum likelihood estimation (MLE) under the corresponding models with the distance measures relating the population and the sample. Their comparison finds that the raking estimator is more efficient than the competitors with a smaller mean squared error for large samples. \cite{meng:rubin93} show that IPF is an expectation/conditional maximization, so the raking estimator shares asymptotic properties similar to MLE and is large-sample Bayes.

We integrate~\eqref{margins-2} with variants of~\eqref{rake-model} into a Bayesian paradigm to incorporate the marginal constraints. Let $\vec{N}_{\cdot\cdot}=(N_{1+},\dots, N_{d_{A}+},N_{+1},\dots, N_{+d_{B}})'$ be the $D (=d_A+d_B)$-length vector of known margins for two variables $A$ and $B$, $\vec{\hat{N}}=(\hat{N}_1,\dots, \hat{N}_{J})'$ be the $J (=d_A \times d_B)$-length vector of cell size estimators, and the loading matrix $L_{D\times J}$ satisfy the constraint of known margins in~\eqref{margins-2}, that is, $L\vec{\hat{N}}=\vec{N}_{\cdot\cdot}$.

If $A$ and $B$ are binary variables with $d_A=d_B=2$, $D=4, J=4$, we have 
\[L=\left(\begin{array}{cccc} 1 & 1 &0 &0\\
0&0&1&1\\
1&0&1&0\\
0&1&0&1
\end{array}\right).\]

To transfer the marginal constraints into Bayesian modeling, we consider two strategies: constructing the solution subspaces and modeling with soft constraints. First, the solution space approach examines the null space of $L$. The dimension of the null space of $L$ is 1, and the basis of Null$(L)$ is $C_{4\times 1}=(1,-1,-1,1)'$. We introduce one free variable $t$ and set an initial estimate $\vec{\hat{N}}_0=(\frac{N_{1\cdot}N_{\cdot1}}{N},\frac{N_{1\cdot}N_{\cdot2}}{N},\frac{N_{2\cdot}N_{\cdot1}}{N},\frac{N_{2\cdot}N_{\cdot2}}{N})'$. 
The basis solution subspace to the linear constraints~\eqref{margins-2} is given by
$
\vec{\hat{N}}=\vec{\hat{N}}_0 + C\vec{t}=(\frac{N_{1\cdot}N_{\cdot1}}{N}+t,\frac{N_{1\cdot}N_{\cdot2}}{N}-t,\frac{N_{2\cdot}N_{\cdot1}}{N}-t,\frac{N_{2\cdot}N_{\cdot2}}{N}+t)'.
$

If we use the projection matrix of the bases C: $P=C(C'C)^{-1}C'$ and assume $\vec{v}=(v_1,v_2,v_3,v_4)'$ is uniformly distributed over the space $R^4$, this provides an alternative estimate with the projection solution subspace 
$
\hat{\vec{N}}=\vec{N}_0 + P\vec{v}=(\frac{N_{1\cdot}N_{\cdot1}}{N}+\frac{v_1-v_2-v_3+v_4}{4},\frac{N_{1\cdot}N_{\cdot2}}{N}-\frac{v_1-v_2-v_3+v_4}{4},
\frac{N_{2\cdot}N_{\cdot1}}{N}-\frac{v_1-v_2-v_3+v_4}{4},\frac{N_{2\cdot}N_{\cdot2}}{N}+\frac{v_1-v_2-v_3+v_4}{4})'.
$

The second method assumes that the margins are drawn from models, such as $\vec{N}_{\cdot\cdot}\sim \textrm{Poisson}(L\vec{N})$, where the constraints are treated as soft and incorporated via modeling mean structure but allow for variation since often the margins are only known approximately.

We propose an integrated procedure under MRP that incorporates the marginal constraints as a unified framework that stabilizes the overall population and domain inferences and accounts for design and response mechanisms. The model-based inference with Bayes-raking has three main components: 1) the margin model incorporating the known constraints; 2) the model for inclusion probability; and 3) the model to predict survey outcome. For large samples, Bayes-raking yields MLE that is equivalent to the raking estimator; for small samples, Bayes-raking is superior to raking with stability and efficiency by smoothing across the hierarchical structure. We describe the general notation for Bayes-raking in the following.

\subsection{The margin model}
\label{margin}

Here we provide a summary of the proposed methods to incorporate the margins into modeling, and the details are provided in the Appendix~\ref{margin}.
The marginal constraints are
\begin{align}
\label{margins-q2}
L\vec{\hat{N}}=\vec{N}_{\cdot\cdot}.
\end{align}

The null space of the matrix $L$ is defined as $\textrm{Null}(L)=\{\vec{N}_L|\vec{N}_L\in R^{J\times 1}, L\vec{N}_L=\vec{0}\}$ with dimension $d_{null}$. Suppose a basis of $\textrm{Null}(L)$ is $C_{J\times d_{null}}$, introduce a vector of free variables $\vec{t}_{d_{null}\times1}=(t_1,\dots,t_{d_{null}})'$, and we have $\vec{N}_L=C\vec{t}$.

The basis solution subspace to the linear constraints~\eqref{margins-q2} is defined as
\begin{align}
\label{subspace}
\{\vec{\hat{N}}|\vec{\hat{N}}=\vec{\hat{N}}_0 + C\vec{t}, \mbox{ } L\vec{\hat{N}}_0=\vec{N}_{\cdot\cdot}, \mbox{ } \vec{t}\in R^{d_{null}\times1},\mbox{ } \vec{N}_0 + C\vec{t}\geq 0\},
\end{align}
 where $\vec{\hat{N}}_0$ is an initial estimator for $\hat{\vec{N}}$ and can be the raking estimate.

The projection solution subspace is defined as
\begin{align}
\label{subspace2}
\{\vec{\hat{N}}|\vec{\hat{N}}=\vec{\hat{N}}_0 + P\vec{v},\mbox{ } L\vec{\hat{N}}_0=\vec{N}_{\cdot\cdot}, \mbox{ } \vec{v}\in R^{J\times1},\mbox{ } \vec{N}_0 + P\vec{v}\geq 0\},
\end{align}
where $P=C(C'C)^{-1}C'$ is the projection matrix of the basis $C$, and $\vec{v}$ is uniformly distributed over the space $R^{J}$. We find that the estimator \eqref{subspace2} is the same as that proposed by~\cite{lazar2008}. Because the number of free variables of $\vec{v}$ in the projection solution subspace is larger than that in the basis solution subspace, that is, $J > d_{null}$, even though the linear combination $C\vec{v}$ is used in the estimate, the computation of the projection solution subspace becomes more difficult.
 
When $d_{null}$ is small, the subspace approaches yield stable inferences under noninformative prior distributions, as demonstrated in our simulation studies. However, the subspace approach can suffer from convergence problem due to the randomness of massive free variables $\vec{t}$ with a large number of sparse or empty cells. It is nontrivial to adapt the generation of $\vec{t}$. While prior distributions could be introduced to $\vec{t}$, the convergence is hard to achieve and calls for structural regularization. In our experiment, we assign the horseshoe distribution~\citep{horseshoe10} as the prior to $\vec{t}$ but fail to achieve convergence even with a modest number of raking variables. This motivates us to propose the second strategy where the marginal constraints in~\eqref{margins-q2} can be softened as model realizations
\begin{align}
\label{model-M}
    \vec{N}_{\cdot\cdot}\sim \textrm{Poisson}(L\vec{N}),
\end{align}
where the marginal constraints are treated as random variables following a Poisson distribution as a natural choice for the counts. Alternative choices include different variance specifications under different distributions. Comparing Poisson distributions with normal distributions with different variance assumptions, our simulation and empirical outputs are not sensitive about the distribution or variance specification in terms of estimation of the population cell counts and the finite population and domain inference.

For empty cells, we introduce a weakly informative prior distribution on $N_j$, $\textrm{Cauchy}^{+}(10,3)$, a half Cauchy distribution with mean 10 and standard deviation 3 restricted to positive values. The modestly small mean value $10$ is chosen because small cells in the population tend to generate empty cells in the sample, and the modestly large scale $3$ allows for flat and heavy tails in the Cauchy distribution~\citep{gelman2008weakly}. The outputs are invariant to the hyperparameter choices in the weakly informative prior distribution setting. The prior belief on the population cell size proportion can be straightforwardly induced in~\eqref{model-M}, such as a loglinear or Dirichlet distribution.

Transferring marginal constraints into prior modeling---either via the solution subspaces~\eqref{subspace}-~\eqref{subspace2} or soft constraints~\eqref{model-M}---will provide posterior estimates for the population cell counts $N_j$'s. We normalize the estimates of $\hat{N}_j$ to achieve the fixed population size $\sum\hat{N}_j=N$. 

\subsection{The sample inclusion model}
\label{selection}

The poststratification cell construction implicitly assumes that the units in each cell are included with equal probabilities. Conditional on fixed sample size $n$,~\cite{BNFP:SI15} consider
%\begin{align}
%\label{multinomial}
\[(n_1,\dots, n_J) \sim \textrm{Multinomial}(n; \frac{N_1p_1}{\sum_j N_jp_j},\dots,\frac{N_Jp_J}{\sum_j N_jp_j}),\]
%\end{align}
where $p_j$ denotes the selection probability for cell $j$. With the goal to estimate domain means based on a large contingency table, this can be approximated by
\begin{align}
\label{selection-n}
n_j \sim \textrm{Poisson}(\frac{nN_jp_j}{\sum_j N_jp_j}).
\end{align}
Here the Poisson distribution can facilitate model fitting via Stan. Assume
\begin{align}
\label{selection-p}
    \textrm{logit}(p_j)  = \alpha_0 + \sum_{k=1}^{K}\sum_{c_k=2}^{d_k}\alpha_{X_{k c_k}^{(j)}}X^{(j)}_{k c_k}.
\end{align}
Here $\alpha_{X_{k c_k}^{(j)}}$ is the coefficient for the indicator $X^{(j)}_{k c_k}$. Without loss of generality, the first level of each raking variable is set as the reference level. The specified model only includes the main effects of the raking variables, the same as the raking model~\eqref{rake-model}, but differs by using the {\em logit} link, rather than the {\em log} link. Our analysis results are similar under the two link choices. The {\em logit} link is an intuitive choice to model the binary sample indicator $I$ and facilitates the Stan computation. Because of known margins, the inclusion model can only be identified from the sample data when main effects are included. When the number of raking variables $K$ or the levels of raking variable $d_k$ is large, informative prior distributions can be elicited to regularize the estimation, and multilevel models can be fit to smooth the estimates. 

The informative prior distribution in Bayes-raking increases the compatibility of including high-order interaction terms among the raking variables that may affect the selection probabilities. Under raking, the correlations are determined by the sample structure while the marginal constraints do not provide extra information. Hence, the classical raking model does not allow the inclusion of high-order interactions. If substantive knowledge shows the true selection probability model depends on high-order interaction terms, we can add such interactions in Bayes-raking and induce proper prior information for model identification. Meanwhile, if the included high-order interaction terms are nonpredictive, Bayes-raking under informative prior distributions will shrink the corresponding coefficients toward 0.

\subsection{The outcome model} 
\label{outcome}

Inside each poststratification cell, the units are identically and independently distributed. For simplicity, when the outcome $y_i$ is continuous, assume $y_i\sim N(\theta_{j[i]},\sigma^2)$; for binary $y_i$, consider $\textrm{logit(Pr}(y_i=1))=\theta_{j[i]}$, where $j[i]$ denotes cell $j$ that unit $i$ belongs to and $\theta_j\sim f(X^{(j)}, \beta_{\theta},\sigma_{\theta})$ with a multilevel model specification. Besides the $X$ variables that affect inclusion, we can add other variables that are predictive for the outcome. Ordinary linear or logistic regression models have $\theta_j=X^{(j)}\beta$. We recommend flexible modeling strategies to predict survey outcome that are robust against misspecification and can stabilize the inference. \cite{BNFP:SI15} introduce Gaussian process as the prior distributions for the functional mean structure under MRP. Alternatives include penalized splines models~\citep{zhenglittle03,zhenglittle05,qchen10} that perform robustly with efficiency gains comparing with classical weighted inference. Under MRP, the cell estimate is a weighted average between the sample cell mean and the overall mean, as smoothed predictions.
 
 \subsection{The unified framework}
 \label{all}
The integrated Bayes-raking procedure systematically induces the marginal constraints as prior distributions, jointly models the survey outcome and the inclusion indicator as the likelihood. For an example with a continuous outcome, the model is summarized as
\begin{align*}
&\textrm{I) Prior: margins}
\begin{cases}
\textrm{1. Basis solution Space:} \\
\mbox{ }\{\vec{\hat{N}}|\vec{\hat{N}}=\vec{\hat{N}}_0 + C\vec{t}, \mbox{ } L\vec{\hat{N}}_0=\vec{N}_{\cdot\cdot}, \mbox{ } \vec{t}\in R^{d_{null}\times1},\mbox{ } \vec{N}_0 + C\vec{t}\geq 0\}\\
\textrm{2. Projection solution Space:}\\
 \mbox{ }\{\vec{\hat{N}}|\vec{\hat{N}}=\vec{\hat{N}}_0 + P\vec{v},\mbox{ } L\vec{\hat{N}}_0=\vec{N}_{\cdot\cdot}, \mbox{ } \vec{v}\in R^{J\times1},\mbox{ } \vec{N}_0 + P\vec{v}\geq 0\}\\
\textrm{3. Soft constraints:} \\
\mbox{ } \vec{N}_{\cdot\cdot}\sim \textrm{Poisson}(L\vec{N})\mbox{, }N_j\sim \textrm{Cauchy}^{+}(10,3) \mbox{ for empty cell } j; \\
\end{cases}\\
&\textrm{II) Likelihood: inclusion mechanism}
\begin{cases}
n_j \sim \textrm{Poisson}(\frac{nN_jp_j}{\sum_j N_jp_j})\\
 \textrm{logit}(p_j)  = \alpha_0 + \sum_{k=1}^{K}\sum_{c_k=2}^{d_k}\alpha_{X_{k c_k}^{(j)}}X^{(j)}_{k c_k};\\
\end{cases}\\
&\textrm{III) Likelihood: survey outcome:}
\begin{cases}
y_i\sim N(\theta_{j[i]},\sigma^2)\\
\theta_j\sim f(X^{(j)}, \beta_{\theta},\sigma_{\theta}).\\
\end{cases}
\end{align*}

Bayes-raking distinguishes from classical raking in five main aspects: 1) Bayes-raking assumes a Bayesian model for the known margins and introduces uncertainty, while raking uses IPF to incorporate the constraint; 2) Bayes-raking treats the sample cell sizes as random variables, which are fixed by raking, and models the cell inclusion probabilities; 3) Bayes-raking uses {\em logit} link while raking uses {\em log} link, resulting in different distance measures; 4) Bayes-raking uses predicted outcome values after smoothing in the MRP estimator, while raking uses the observed sample values; and 5) Bayes-raking predicts population cell counts and estimates for empty cells that are omitted by raking.

This open-ended framework can accommodate additional hierarchical structures. The three components of the unified inference framework are fitted in Stan as an integrated and fully Bayesian inference. The computation via Stan is efficient with a user-friendly interface. The codes are deposited in GitHub~\citep{rake:code19} and ready for public use.

%%%%%%%%%%%%%%%%%%%%%%%%%%%%
\section{Simulation studies}
\label{simulation}

We use simulation studies to evaluate the Bayes-raking estimation and compare with IPF. We treat the 2011 American Community Survey (ACS) survey of $51714$ New York City (NYC) adult residents as the ``population" and perform repeated sampling to check the statistical validity. We randomly draw 150 repetitions out of the ACS according to a pre-specified inclusion model. We simulate the outcome variable in the ACS as the truth. We implement the three Bayes-raking approaches depending on the margin model choices: the projection solution subspace using the projection matrix that is similar to \cite{lazar2008}; the basis solution subspace using the basis functions; and the Bayes-model with soft modeling constraints. 

Focusing on the overall population and domain mean estimates, we calibrate the Bayes-raking estimation by examining the bias, root mean squared error (RMSE), standard error (SE) approximated by the average of standard deviations and nominal coverage rate (CR) of the 95\% confidence intervals. When no empty cells occur, our experiments show that the Bayesian approaches perform as well as raking with similar outputs. Hence, we present the cases with empty cells when the theoretical properties of IPF are not satisfied: the scenarios with a sparse contingency table and dependent raking variables.

\subsection{Sparse contingency table}

Survey practice often selects a modest number of raking variables, resulting in a sparse contingency table. The sparsity will be severe with an increasing number of raking variables. As an illustration, we collect five raking variables in the 2011 ACS data: age ({\em age}: 18--34, 35--44, 45--54, 55--64, 65+), race/ethnicity ({\em race}: white \& non-Hispanic, black \& non-Hispanic, Asian, Hispanic, other), education ({\em edu}: less than high school, high school, some college, bachelor degree or above), sex, and poverty gap\footnote{Poverty gap index is a measure of the intensity of poverty and defined as a proportion of the poverty line.} ({\em pov}: under 50\%, 50-100\%, 100-200\%, 200-300\%, 300\%+; Higher {\em pov} value means wealthier). The cross-tabulation of the five ACS raking variables constructs 1000 ($5\times 5\times 4 \times 2 \times 5$) poststratification cells, 14 cells have no units, and the $J=986$ nonempty cells sizes vary between 1 and 1283. 

Assume the outcome is binary and generated via a logistic regression: $\textrm{logit(Pr(}y_i=1))=X_i\beta$, where $X_i$ represents the main effects for the five ACS raking variables of unit $i$. The nonzero components of the coefficient vector $\beta$ are $(\beta_{0}=0.85, \beta_{age,2}=0.41, \beta_{age,3}=0.48, \beta_{age,4}=0, \beta_{age,5}=-0.63, \beta_{race,2}=1, \beta_{race,3}=0, \beta_{race,4}=1.14, \beta_{race,5}=1.28, \beta_{edu,2}=0, \beta_{edu,3}=0, \beta_{edu,4}=-0.81, \beta_{fem}=0.31, \beta_{pov,2}=-0.61, \beta_{pov,3}=0, \beta_{pov,4}=-0.78, \beta_{pov,5}=-1.38)$. We consider the sample inclusion model without nonresponse: $\textrm{logit}(p_i)=X_i\alpha$, where $p_i$ is the selection probability of unit $i$. The coefficient vector $\alpha$ has nonzero components: $(\alpha_{0}=-4.31, \alpha_{age,2}=0.26, \alpha_{age,3}=0.46, \alpha_{age,4}=0.57, \alpha_{age,5}=0.51, \alpha_{race,2}=0.43, \alpha_{race,3}=-0.84, \alpha_{race,4}=1.13, \alpha_{race,5}=0.68, \alpha_{edu,2}=0.47, \alpha_{edu,3}=0.64, \alpha_{edu,4}=1.19, \alpha_{fem}=0.32, \alpha_{pov,2}=0, \alpha_{pov,3}=-0.26, \alpha_{pov,4}=-0.46, \alpha_{pov,5}=-0.48)$. The values of $\alpha$ and $\beta$ are chosen to mimic the exploratory analysis structure of the LSW study.

Assume the marginal distributions of the five raking variables are known, but not the joint distribution. We are interested in the overall population mean estimation, the subgroup means of marginal levels for each raking variable, and the subgroup means of the two-way interaction between age and poverty gap. We implement the three Bayes-raking estimation approaches and IPF. The two subspace approaches cannot converge with 986 cells because of computational difficulty with the high dimensional free variables. Therefore, we only report the outputs of Bayes-raking with the soft Poisson modeling constraint and IPF. 

To emphasize the comparison between Bayes-raking and raking on their capability of accounting for design effect and calibration, we set the fitted outcome model the same as the data generation model. The true model parameters can be recovered via Stan. In practice, we recommend flexible models with diagnostics and evaluation as Section~\ref{application}. We repeatedly draw samples from the ACS data with sample sizes around 2000, where the number of empty cells is between 368 and 424 across repetitions. 

\begin{figure}
    \includegraphics[width = \linewidth]{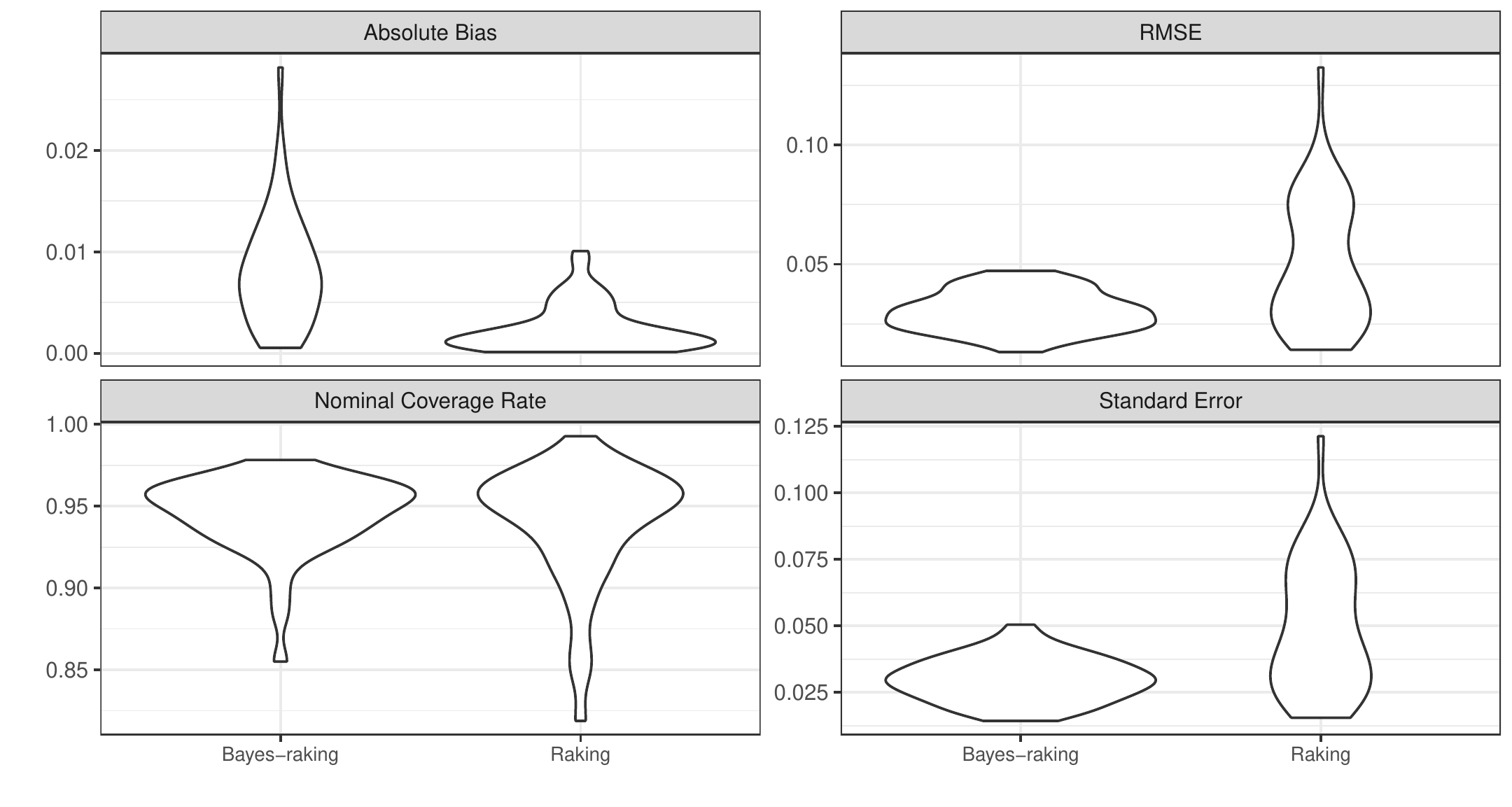}
    \caption{Comparison on domain mean estimation for 46 subgroups defined by the marginal levels of five raking variables: age, race, education, sex and poverty gap, and the interaction between age and poverty gap. Bayes-raking yields smaller RMSE and SE with more reasonable nominal coverage rates than raking, particularly for the subgroups defined by the interaction.}
    \label{sim1}
\end{figure}

For the overall population mean, both methods give similar results, with the same mean estimate (0.59), SE (0.01) and RMSE (0.01) and similar coverage rates, 0.94 under Bayes-raking and 0.96 under raking. When looking at the subgroup mean estimates shown in Figure~\ref{sim1}, Bayes-raking outperforms raking. Bayes-raking and raking generate similar estimates that are generally unbiased though raking has negligibly smaller biases. Bayes-raking yields smaller RMSE and SE with more reasonable nominal coverage rates than raking, particularly for the subgroups defined by the interaction. Raking has low coverage rates for the domain mean estimates of the groups who are older than 35 with low income (poverty gap below 50\%), where the coverage rates are below 0.90 (0.82, 0.86, 0.89, 0.87). These groups have relatively small population cell sizes, between 330 and 586. Raking fails to yield valid inferences with small cell sizes. With empty cells in the sample, the theoretical guarantee for IPF convergence is not satisfied. Bayes-raking performs stably by treating the sample cell sizes as random variables generated from the Poisson models with the inclusion probability models.

\subsection{Dependent raking variables}

To examine the performances of Bayes-raking and raking with dependent raking variables, we collect three ACS variables: age, poverty gap and the number of children in the family ($cld: 0, 1, 2, 3+$) and use the two-way interaction terms in the raking adjustment: $age\cdot pov$ and $cld \cdot pov$. This scenario is motivated by the substantive analysis questions of interest, such as the percentage of elders or children living in poverty. 

Both two-way interaction terms have $pov$ and are correlated, with a high $Cram\acute{e}r's$ $V$ value $0.51$. However, IPF treats the interactions as independent in a two-dimension contingency table ($25\times 20$). The proposed Bayes-raking estimation accounts for the cross-classified structure and thus the correlation between the two interaction terms.

The three variables construct 100 $(=5\times 5\times 4)$ poststratification cells. We treat the ACS data as the population with population cell sizes varying between 12 and 5462 with a median 248. The assumed inclusion model includes main effects and two-order interactions: $\textrm{logit}(p_i)=\alpha_0+age_i\vec{\alpha}_{age}+pov_i\vec{\alpha}_{pov}+cld_i\vec{\alpha}_{cld}+age_i\times pov_i\vec{\alpha}_{age, pov}+cld_i\times pov_i\vec{\alpha}_{cld, pov}$. The nonzero components of the coefficients $\vec{\alpha}$ are $(\alpha_0=-3.45, \alpha_{age,3}=0.53, \alpha_{age,4}=0.32, \alpha_{age,5}=0.76, \alpha_{pov,2}=0.12, \alpha_{pov,5}= -0.30, \alpha_{cld,2}=0.48, \alpha_{cld,3}=0.17, \alpha_{cld,4}=-0.17, \alpha_{age5,pov2}=-0.87,\alpha_{age5,pov5}= - 0.30, \alpha_{pov2,cld2}=-0.52, \alpha_{pov5,cld4}=0.46)$. Across 150 repeatedly selected samples, the sample sizes are centered around 2000 with the number of empty cells close to 5, which tend to be middle-aged (45-54) or elder (65+) individuals with low income (pov$<$50\%) and many children (3+). The sample cell sizes are between 1 and 127 with a median 10. 

We consider the outcome model: $y_{i}=1/p_{i}+\epsilon_i$, $\epsilon_i\sim N(0,1/p^2_i)$, the implicit model for the Horvitz-Thompson estimator~\citep{ht52}, to introduce high correlation between the outcome and the selection probability. When the selection probability is small, the outcome has a large mean and high variance. The estimation model is the same as the data generation model. We compare the three Bayes-raking estimation approaches and IPF.

Table~\ref{sim2-overall} displays the comparison for the overall mean estimation, where the four methods perform similarly well. IPF generates negligibly smaller bias but larger SE and RMSE than the Bayesian approaches. Because of the wide confidence intervals, IPF has more conservative coverage than the Bayesian approaches, the coverage rates of all which are reasonable.

\begin{table}[t]
\centering
\caption{Outputs on overall population mean estimation. Projection denotes Bayes-raking estimation with projection solution subspace; Basis represents Bayes-raking estimation with the basis solution subspace that we proposed; and Bayes-model refers to Bayes-raking estimation with soft modeling constraints. The four methods perform similarly well.}
\label{sim2-overall}
\begin{tabular}{lrrrrrr}
\hline
& Mean  & Bias   & RMSE    &SE &  $95\%$ CI length & Coverage  \\ 
\hline
Projection     & 28.71 & 0.23  & 0.57 & 0.52 & 2.04     & 0.94                 \\
Basis  & 28.71   & 0.23  & 0.58 & 0.52& 2.02     & 0.93                 \\
Bayes-model& 28.75& 0.26  & 0.57  & 0.55 & 2.17     & 0.94                  \\
Raking     & 28.48  & 0.00 & 0.70 & 0.76& 2.98     & 0.97  \\
\hline              
\end{tabular}
\end{table}

\begin{figure}[ht!]
    \begin{tabular}{cc}
        \includegraphics[width = 0.5\linewidth]{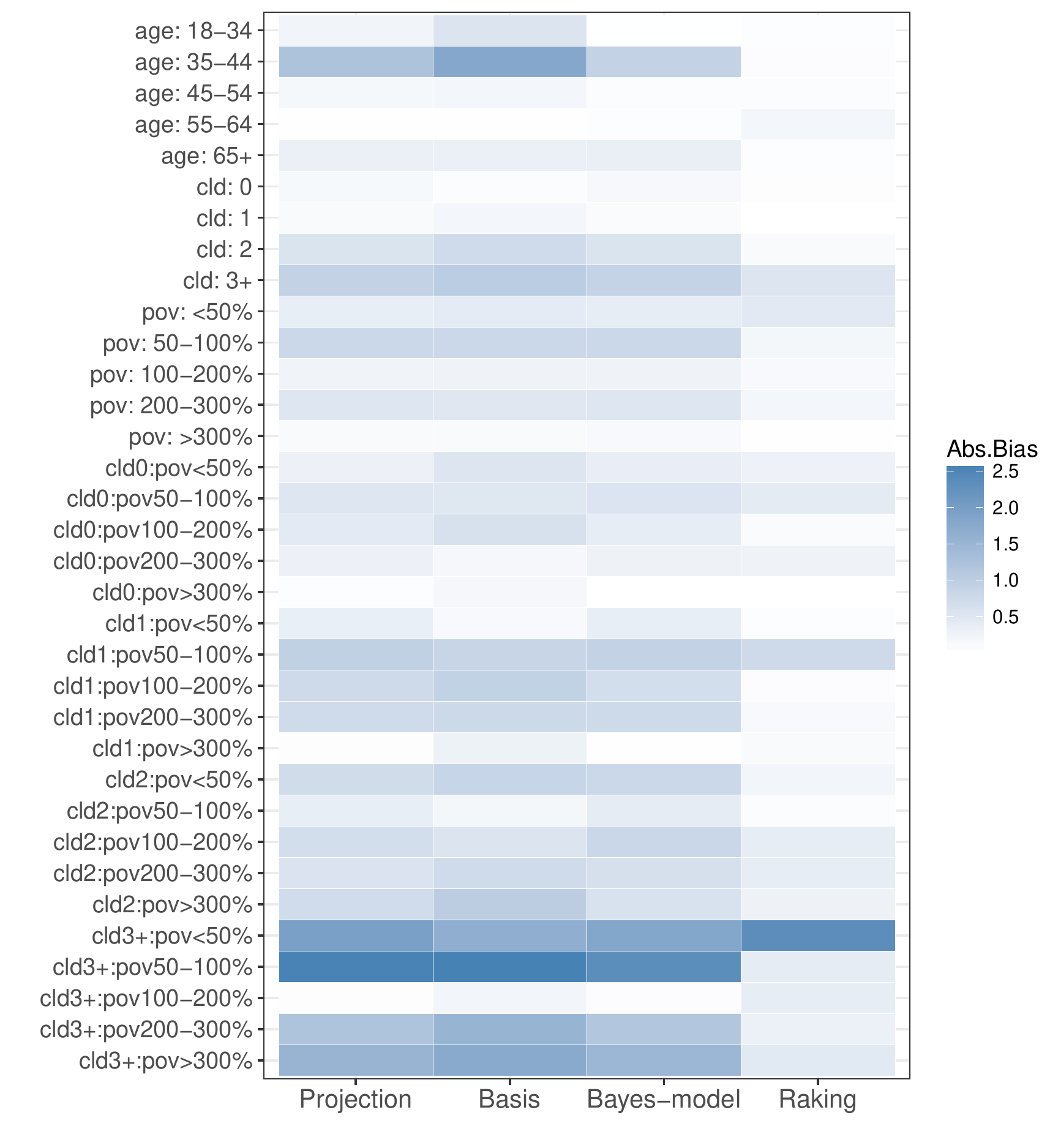}
        \includegraphics[width = 0.5\linewidth]{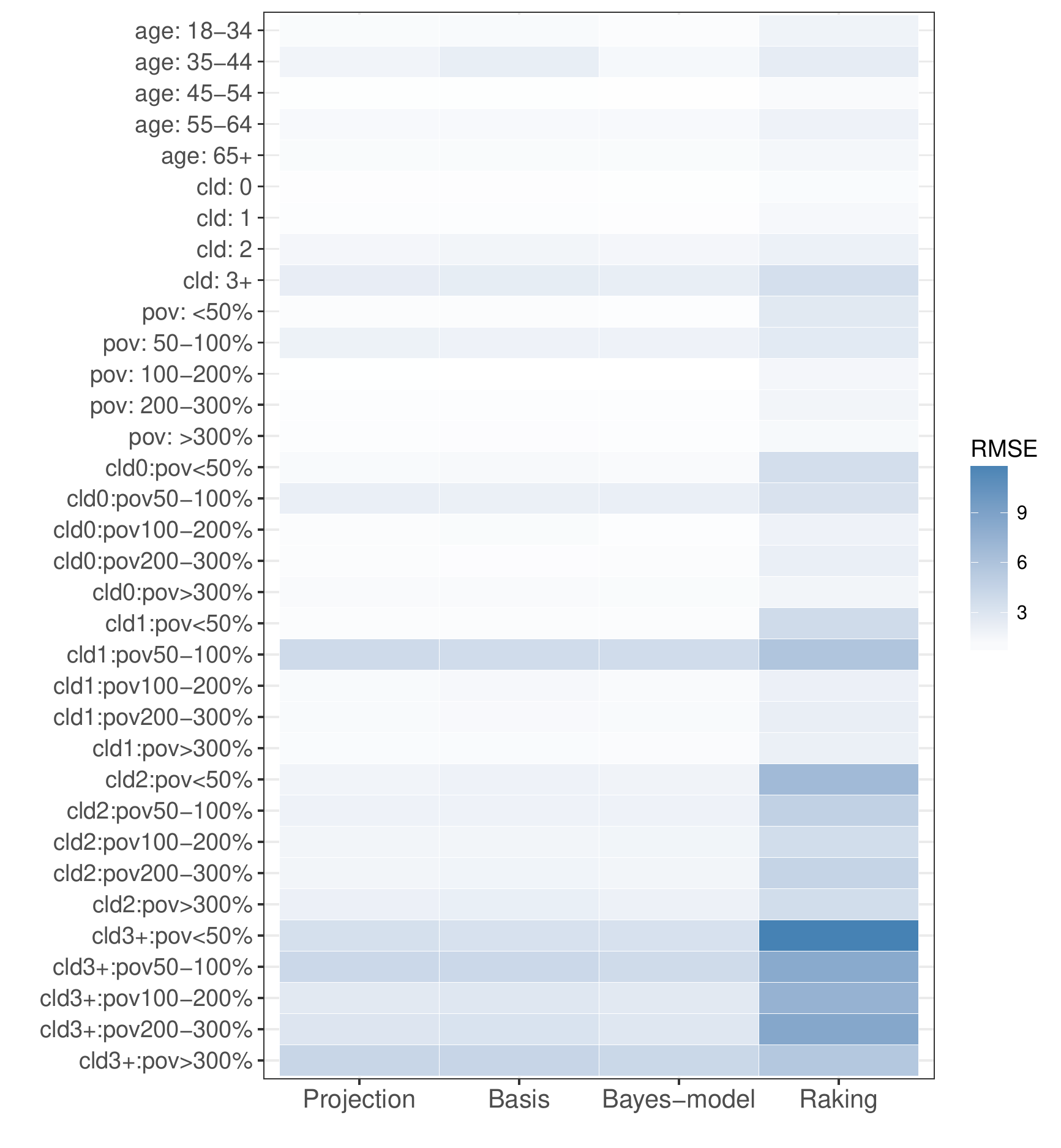} \\
        \includegraphics[width = 0.5\linewidth]{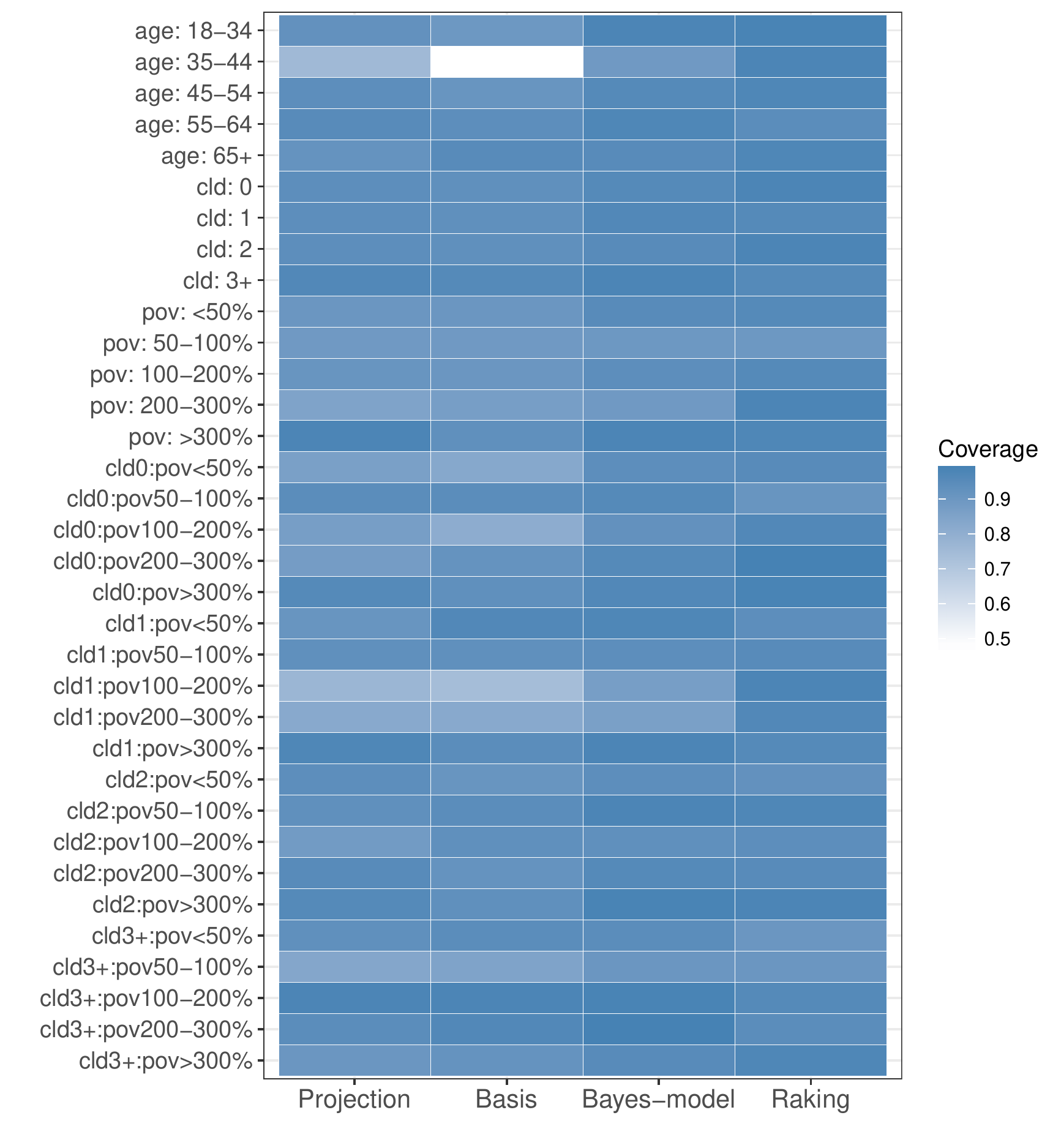}
        \includegraphics[width = 0.5\linewidth]{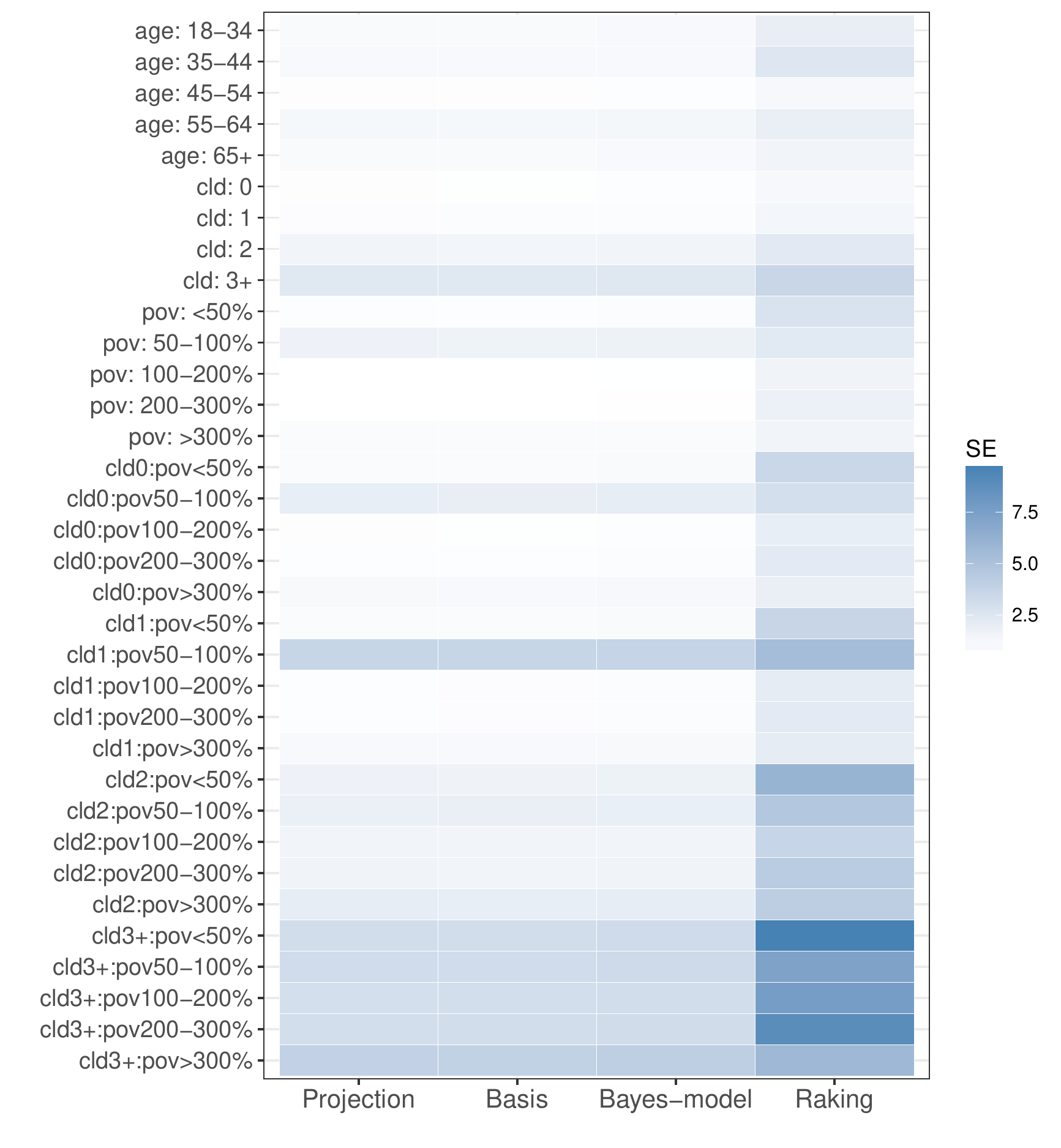}
    \end{tabular}
    \caption{Outputs on finite domain mean estimation for subgroups defined by the marginal levels of three variables: age, poverty gap and number of children in the family ($cld$), and subgroups defined by the interaction between the number of children in the family and poverty gap. Projection denotes Bayes-raking estimation with projection solution subspace; Basis represents Bayes-raking estimation with the basis solution subspace that we proposed; and Bayes-model refers to Bayes-raking estimation with soft modeling constraints. }
    \label{sim2}
\end{figure}

The efficiency gains of Bayesian approaches are further demonstrated in the domain inference in Figure~\ref{sim2}, where the IPF algorithm yields substantially larger SE and RMSE than the Bayesian approaches, $\sim 3$ times, even with competitive bias and coverage rates. The correlation between the two raking factors makes the algorithm hard to converge and yields highly variable inference. The Bayesian approaches reconcile the hierarchy structure with the high-order interactions and are capable of incorporating the known margin constraints while propagating the uncertainty. 

Among the three Bayesian approaches, the performances are generally similar in terms of SE and RMSE. The Bayes-raking estimation with modeling yields more reasonable coverage rates than the subspace approaches that could yield low coverage rates. The subspace approaches have large bias and low coverage rate when estimating the domain mean for those whose age is between 35 and 44. This could be related to the computational performance of the subspace approaches, where the convergence is sensitive about the solution path and hard to be guaranteed during the repeated sampling process. We generally recommend the Bayes-raking estimation with soft modeling constraints that perform stably under a large number of raking variables with dependency.

%%%%%%%%%%%%%%%%%%%%%%%%%%%%
\section{LSW study application}
\label{application}

%LSW
The methodology development is motivated by our survey weighting practice for the NYC Longitudinal Study of Wellbeing (LSW) \citep{RHreport}, which is a biennial survey (2013--2014) aiming to provide assessments of income poverty, material hardship, and child and family wellbeing of NYC adult residents. The sample includes a phone sample based on random digit dialing and an in-person respondent-driven sample of beneficiaries from Robin Hood philanthropic services and their acquaintances. We focus on the phone survey here as an illustration. The LSW phone survey interviews 2002 NYC adult residents based on 500 cell phone calls and 1502 landline telephone calls, where half of the landline samples are from poor NYC areas defined by zip code information. These collected baseline samples are followed up for every three months. The sample is not representative of the target population due to the oversampling design and differential inclusion propensities. To balance the discrepancies, we match the samples to the 2011 ACS for NYC adult records. 

The baseline weighting process~\citep{RHweighting} adjusts for the unequal probability of selection, nonresponse and coverage bias. The selection of variables for weighting is subjective depending on the conventional weighting practice routine and the organizers' analysis of interest, for example, how many NYC children living in poverty. Raking is implemented by matching the marginal distributions of the selected factors. We apply Bayes-raking to the LSW study. The poststratification weighting factors include sex, age, education, race, poverty gap, the number of children in the family, the two-way interaction between age and poverty gap, and the two-way interaction between the number of children in the family and poverty gap. We collect the marginal distributions of the poststratification factors from ACS\footnote{ACS has its own weights, and the marginal distributions have accounted such weights to be representative of the population.} and implement the two raking methods. Similar to the cases in Section~\ref{simulation}, some raking variables are correlated. IPF treats these factors independently, while Bayes-raking procedure accounts for the hierarchy structure of high-order interactions.

\begin{table}[ht]
\centering
\caption{Proportion estimation for NYC residents suffering from material hardship with standard errors reported in parenthesis.}
\label{lsw1}
\begin{tabular}{lll}
  \hline
& Bayes-raking & Raking  \\ 
  \hline
Overall & 0.607 (0.012) & 0.604 (0.012) \\ 
  Age: 18-34 & 0.613 (0.022) & 0.622 (0.025) \\ 
  Age: 35-44 & 0.645 (0.024) & 0.651 (0.027) \\ 
  Age: 45-54 & 0.675 (0.024) & 0.667 (0.027) \\ 
  Age: 55-64 & 0.584 (0.023) & 0.569 (0.028) \\ 
  Age: 65+ & 0.48 (0.025) & 0.473 (0.027) \\ 
 Less than high sch& 0.742 (0.022) & 0.799 (0.028) \\ 
High sch & 0.711 (0.021) & 0.711 (0.026) \\ 
Some coll & 0.674 (0.021) & 0.669 (0.028) \\ 
Bachelor or above & 0.394 (0.017) & 0.361 (0.021) \\ 
Male & 0.555 (0.019) & 0.555 (0.021) \\ 
Female & 0.648 (0.014) & 0.647 (0.015) \\ 
  Pov: $<$50\% & 0.785 (0.025) & 0.859 (0.031) \\ 
  Pov: 50-100\% & 0.763 (0.024) & 0.782 (0.03) \\ 
  Pov: 100-200\% & 0.756 (0.021) & 0.747 (0.029) \\ 
  Pov: 200-300\% & 0.684 (0.03) & 0.705 (0.033) \\ 
  Pov: $>$300\% & 0.453 (0.016) & 0.434 (0.019) \\ 
White\&non-Hisp& 0.424 (0.019) & 0.417 (0.022) \\ 
 Black\&non-Hisp& 0.739 (0.018) & 0.737 (0.021) \\ 
Asian & 0.444 (0.035) & 0.419 (0.055) \\ 
Hisp & 0.727 (0.041) & 0.729 (0.051) \\ 
Other & 0.803 (0.018) & 0.829 (0.017) \\ 
None cld& 0.553 (0.014) & 0.542 (0.016) \\ 
 \#Cld: 1 & 0.679 (0.026) & 0.688 (0.031) \\ 
 \#Cld: 2 & 0.706 (0.028) & 0.732 (0.029) \\ 
 \#Cld: 3+ & 0.736 (0.037) & 0.766 (0.041) \\ 
   \hline
\end{tabular}
\end{table}

\begin{figure}[ht!]
    \begin{tabular}{c}
        \includegraphics[width =0.95\linewidth]{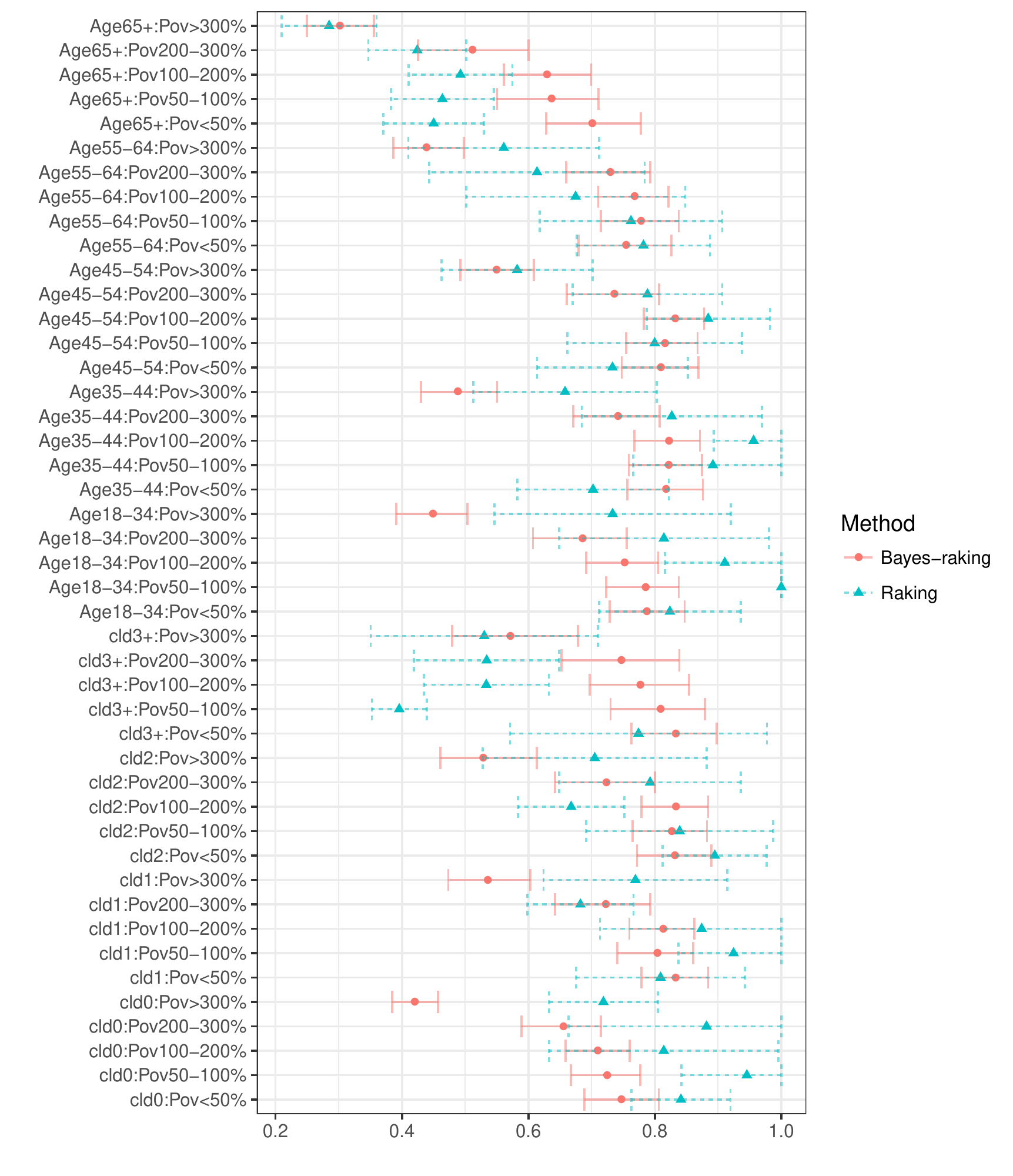}
    \end{tabular}
    \caption{Proportion estimation for subgroups of NYC residents suffering from material hardship. Bayes-raking performs more efficiently with smaller standard errors than raking.}
    \label{lsw}
\end{figure}
The outcome of interest is the material hardship indicator: $Y_i=1$ indicates individual $i$ suffers from material hardship; $Y_i=0$ otherwise. We perform model evaluation and comparison using the leave-one-out cross validation~\citep{loo17}, and the outcome model with only main effects outperforms those with interactions by generating competitive expected log point-wise predictive density with the smallest number of parameters. Hence the selected outcome model includes the main effects of sex, age, education, race, poverty gap and the number of children in the family.

We compare the performance of Bayes-raking and raking on the proportion estimation of those who have material hardship among all NYC adult residents and subgroups of interest. The two subspace approaches in the Bayesian paradigm cannot converge, so we focus on the comparison between Bayesian modeling with soft constraints and IPF. 

Table~\ref{lsw1} and Figure~\ref{lsw} present the proportion estimation for NYC residents suffering from material hardship using the Bayes-raking and raking procedures, respectively. Table~\ref{lsw1} displays the estimation based on all NYC residents and subgroups defined by the marginal levels of the six raking variables. In Table~\ref{lsw1}, while the estimates are similar among the two approaches with overlapping confidence intervals, Bayes-raking generally yields smaller standard error than raking. The similarity is due to the large sample sizes as expected. As illustrated in Figure~\ref{lsw}, the domain inference is more efficient with shorter confidence intervals under Bayesian raking and is significantly different for some groups between the two approaches.

Comparing with classical raking, Bayes-raking generates significantly lower proportions of NYC residents with material hardship for families without any child and poverty gap 50-100\% or above 300\%, families with one child and poverty gap above 300\%, youth (18-34) with poverty gap above 300\%, and early middle-aged (35-44) with poverty gap 100-200\%, and has significantly higher estimates on material hardship proportion for those with two children and poverty gap 100-200\%, families with three children and poverty gap below 300\%, and elder (65+) with poverty gap below 200\%. Briefly, Bayes-raking yields lower proportional estimates for material hardship for young NYC residents with a fewer number of children and higher income, but higher estimates for those with many children, elder, and lower income. The LSW study oversamples poor individuals, and the estimates on material hardship without effective adjustment for sampling or nonresponse bias will be overestimated. Bayes-raking substantively improves by correcting the overestimation and presenting inferential efficiency.

Furthermore, only relying on the collected sample statistics, raking can yield unreasonable estimates and high variability. For youth (18-34) with a poverty gap 50-100\%, raking has a proportion with material hardship at 1 without any variability. However, Raking yields large standard errors for the families with three more children and poverty gap below 50\% and youth (18-34) with poverty gap above 300\%. Bayes-raking borrows information across subgroups to smooth the estimates and hence performs more robustly and efficiently than IPF adjustment.

%%%%%%%%%%%%%%%%%%%%%%%%%%%%
\section{Conclusion}
\label{discussion}

We develop a Bayesian paradigm for raking to incorporate marginal constraints from two directions: utilizing basis and projection solution subspaces and modeling soft constraints. Bayes-raking accounts for the cross-classified structure in a principled way and smoothes cell estimates especially with sparse contingency tables. Bayes-raking is equivalent to raking as large-sample MLE if the theoretical guarantee of IPF is satisfied, and outperforms raking when IPF suffers from convergence issues, such as cases with a sparse contingency table or dependent raking variables. Bayes-raking yields statistical validity and efficiency gains, with the capability of uncertainty propagation. 

We use Bayes-raking to supplement MRP as a model-assisted, design-based approach to a unified framework for survey inference. Simulation experiments and application studies demonstrate the improvement of Bayes-raking on frequentist properties and substantive findings. Furthermore, Bayes-raking increases the weighting compatibility and is ready to be induced with additional prior distributions, which facilitates computation and information fusion. Bayes-raking has great potential to calibrate nonprobability samples, especially the studies with informative sampling schemes and nonignorable nonresponse, and assist data integration with a systematic framework.

The integrated Bayes-raking procedure opens several interesting directions worth further investigation. First, Bayes-raking makes inferences of the population cell counts, while raking minimizes the KL divergence between the estimates and the sample sizes. \cite{littlewu91} study various distance functions, such as the least square estimate, maximum likelihood estimation and estimate minimizing chi-squared error. Bayes-raking uses a logistic regression for the inclusion probability and a Poisson model for the cell size, which could result in a different distance measure. 
It is unclear how the various distance measures are related to design and asymptotic efficiency. Future work is needed to examine the theoretical properties and choose a proper distance measure that can directly access the estimation of population cell sizes as an intermediate inference goal and benefit the ultimate goal of stable finite population and domain inference. This is also related to weight smoothing on how to select proper loss or distance functions as an object to determine the optimal weights~\citep{modelass-sarndal92,prior-si2018}.

Second, the model-based approach uses survey outcome and then has potential efficiency gain. For probability sampling, ideally, the poststratification cells should be constructed using variables that predict both the outcome and selection, where within-cell variability is minimized and between-cell variability is maximized. As a byproduct, the outcome can inform variable selection for raking. The variables that are not predictive of the outcome could be omitted from raking, in particular when raking suffers from convergence issues with sparse cell structure. Bayes-raking can be extended by imposing connections between the coefficients in the outcome model and those in the inclusion model. This shall facilitate computation and inference efficiency, but demand a tradeoff between efficiency and robustness.

Furthermore, we recommend flexible models for the outcome, such as nonparametric Bayesian models, to achieve inferential stability against model misspecification. Further work is needed to invest theoretical properties of the Bayes-raking estimator under MRP, such as double robustness, in comparison with calibration weighting estimators. Extra information about the selection process can be leveraged via prior distributions. In the paper, we focus on selection assuming full participation. While nonresponse is unavoidable, the variables that affect response propensity can be included in the raking process if available. However, informative nonresponse or inclusion mechanism needs further investment. Bayes-raking builds up the foundation of incorporating marginal constraints into modeling and is ready to integrate complex design and response features with flexible outcome modeling as a unified framework for survey inference.

\appendix

\section{Supplemental materials}
\label{margin}
Here we provide the detailed description of the marginal models.

In the basic setting, the $K$ variables $X=(X_1,\dots, X_K)$ that affect sample inclusion propensities are used for raking to balance the sample and the population, and their cross-tabulation constructs poststratification cells. Here we assume $X_k$'s are discrete (after discretization of continuous variables) with possible values $c_k=1,\dots,d_k$, where $d_k$ is the total number of levels. All the possible combination categories of $X$ are labeled as cell $j$, with value $X^{(j)}=(X_1^{(j)},\dots, X_K^{(j)})'$, $X_k^{(j)}=(X_{k1}^{(j)}, \dots, X_{kd_k}^{(j)})'$, where $X_{kc_k}^{(j)}$'s are $0/1$ indicators, the population cell size $N_j$ and the sample cell size $n_j$, for $j=1,\dots, J$, where $J$ is the total number of poststratification cells. The total population size is $N=\sum_{j=1}^J N_j$, and the sample size is $n=\sum_{j=1}^Jn_j$. Denote $D=\sum_kd_k$, and $D\leq J$. The joint distribution of $X$ is unknown, but the marginal distributions are available. That is, the population cell counts $N_j$'s are unknown, and the estimators $\hat{N}_j$'s are subject to marginal constraints:
\begin{align}
\label{ap:margins-q}
\sum_{\{j: X^{(j)}_{k c_k}=1\}} \hat{N}_{j}=N_{kc_k+},
\end{align}
where $N_{kc_k+}$'s denote the known margins for the $c_k$th level of $X_k$, $c_k=1,\dots, d_k, k=1,\dots, K$. Let $\vec{N}_{k\cdot}=(N_{k1+},\dots, N_{kd_k+})'$ be the $d_k\times 1$ vector of known margins for Variable $X_k$, and $\vec{N}_{\cdot\cdot}=(\vec{N}'_{1\cdot},\dots,\vec{N}'_{K\cdot})'$ be the $D\times 1$ vector of known margins for all $X$ variables. Denote $\vec{\hat{N}}=(\hat{N}_1,\dots,\hat{N}_J)'$ as all cell size estimators and introduce a $J\times1$ loading vector $L_{kc_k}$ that maps the marginal constraint: $L'_{kc_k}\vec{\hat{N}}=N_{kc_k+}$.  Denote the $d_k\times J$ loading matrix $L_{k\cdot}=(L'_{k1},\dots, L'_{kd_k})'$, and the $D\times J$ loading matrix $L=(L'_{1\cdot},\dots,L'_{K\cdot})'$. Then we have 
$L_{k\cdot}\vec{\hat{N}}=\vec{N}_{k\cdot}$. The constraints in~\eqref{ap:margins-q} have the equivalent expression as
\begin{align}
\label{ap:margins-q2}
L\vec{\hat{N}}=\vec{N}_{\cdot\cdot}.
\end{align}

The null space of the matrix $L$ is defined as $\textrm{Null}(L)=\{\vec{N}_L|\vec{N}_L\in R^{J\times 1}, L\vec{N}_L=\vec{0}\}$ with dimension $d_{null}$. Suppose a basis of $\textrm{Null}(L)$ is $C_{J\times d_{null}}$, introduce a vector of free variables $\vec{t}_{d_{null}\times1}=(t_1,\dots,t_{d_{null}})'$, and we have $\vec{N}_L=C\vec{t}$.

The basis solution subspace to the linear constraints~\eqref{ap:margins-q2} is defined as
\begin{align}
\label{ap:subspace}
\{\vec{\hat{N}}|\vec{\hat{N}}=\vec{\hat{N}}_0 + C\vec{t}, \mbox{ } L\vec{\hat{N}}_0=\vec{N}_{\cdot\cdot}, \mbox{ } \vec{t}\in R^{d_{null}\times1},\mbox{ } \vec{N}_0 + C\vec{t}\geq 0\},
\end{align}
 where $\vec{\hat{N}}_0$ is an initial estimator for $\hat{\vec{N}}$ and can be the raking estimate.

As an alternative solution, the projection solution subspace is defined as
\begin{align}
\label{ap:subspace2}
\{\vec{\hat{N}}|\vec{\hat{N}}=\vec{\hat{N}}_0 + P\vec{v},\mbox{ } L\vec{\hat{N}}_0=\vec{N}_{\cdot\cdot}, \mbox{ } \vec{v}\in R^{J\times1},\mbox{ } \vec{N}_0 + P\vec{v}\geq 0\},
\end{align}
where $P=C(C'C)^{-1}C'$ is the projection matrix of the basis $C$, and $\vec{v}$ is uniformly distributed over the space $R^{J}$. 
%%%%%%%%%%%%%%%%%%%%%%%%%%%%
\bibliography{weighting-Jun2019}

\begin{thebibliography}{}

\bibitem[\protect\citeauthoryear{Battaglia, Hoaglin, and Frankel}{Battaglia
  et~al.}{2013}]{battaglia2013practical}
Battaglia, M.~P., D.~C. Hoaglin, and M.~R. Frankel (2013).
\newblock Practical considerations in raking survey data.
\newblock {\em Survey Practice\/}~{\em 2\/}(5), 1--10.

\bibitem[\protect\citeauthoryear{Bishop and Fienberg}{Bishop and
  Fienberg}{1969}]{bishop1969incomplete}
Bishop, Y.~M. and S.~E. Fienberg (1969).
\newblock Incomplete two-dimensional contingency tables.
\newblock {\em Biometrics\/}~{\em 25}, 119--128.

\bibitem[\protect\citeauthoryear{Bishop, Fienberg, and Holland}{Bishop
  et~al.}{1974}]{bishop75}
Bishop, Y.~M., S.~E. Fienberg, and P.~W. Holland (1974).
\newblock {\em Discrete Multivariate Analysis: Theory and Practice}.
\newblock Springer.

\bibitem[\protect\citeauthoryear{Breidt and Opsomer}{Breidt and
  Opsomer}{2017}]{model-assist-review-SS17}
Breidt, F. and J.~Opsomer (2017).
\newblock Model-assisted survey estimation with modern prediction techniques.
\newblock {\em Statistical Science\/}~{\em 32}, 190--205.

\bibitem[\protect\citeauthoryear{Brick, Montaquila, and Roth}{Brick
  et~al.}{2003}]{raking:brick03}
Brick, J.~M., J.~Montaquila, and S.~Roth (2003).
\newblock Identifying problems with raking estimators.
\newblock 2003 Joint Statistical Meetings, Section on Survey Research Methods.

\bibitem[\protect\citeauthoryear{Carvalho, Polson, and Scott}{Carvalho
  et~al.}{2010}]{horseshoe10}
Carvalho, C.~M., N.~G. Polson, and J.~G. Scott (2010).
\newblock The horseshoe estimator for sparse signals.
\newblock {\em Biometrika\/}~{\em 97}, 465--480.

\bibitem[\protect\citeauthoryear{Chen, Elliott, and Little}{Chen
  et~al.}{2010}]{qchen10}
Chen, Q., M.~R. Elliott, and R.~J. Little (2010).
\newblock Bayesian penalized spline model-based inference for finite population
  proportion in unequal probability sampling.
\newblock {\em Survey Methodology\/}~{\em 36\/}(1), 23--34.

\bibitem[\protect\citeauthoryear{Deming and Stephan}{Deming and
  Stephan}{1940}]{deming1940least}
Deming, W.~E. and F.~F. Stephan (1940).
\newblock On a least squares adjustment of a sampled frequency table when the
  expected marginal totals are known.
\newblock {\em The Annals of Mathematical Statistics\/}~{\em 11\/}(4),
  427--444.

\bibitem[\protect\citeauthoryear{Fienberg}{Fienberg}{1970}]{fienberg1970iterative}
Fienberg, S.~E. (1970).
\newblock An iterative procedure for estimation in contingency tables.
\newblock {\em The Annals of Mathematical Statistics\/}~{\em 41\/}(3),
  907--917.

\bibitem[\protect\citeauthoryear{Folsom and Singh}{Folsom and
  Singh}{2000}]{gem:Folsom}
Folsom, R.~E. and A.~Singh (2000).
\newblock The generalized exponetial model for sampling weight calibration for
  extreme values, nonresponse, and poststratification.
\newblock Proceedings of the Survey Research Methods Section, American
  Statistical Association.

\bibitem[\protect\citeauthoryear{Gelman and Carlin}{Gelman and
  Carlin}{2001}]{gelmancarlin01}
Gelman, A. and J.~B. Carlin (2001).
\newblock Poststratification and weighting adjustments.
\newblock In R.~Groves, D.~Dillman, J.~Eltinge, and R.~Little (Eds.), {\em
  Survey Nonresponse}.

\bibitem[\protect\citeauthoryear{Gelman, Carlin, Stern, Dunson, Vehtari, and
  Rubin}{Gelman et~al.}{2014}]{gelman14bda}
Gelman, A., J.~B. Carlin, H.~S. Stern, D.~Dunson, A.~Vehtari, and D.~B. Rubin
  (2014).
\newblock {\em Bayesian Data Analysis\/} (3nd ed.).
\newblock CRC Press, London.

\bibitem[\protect\citeauthoryear{Gelman, Carlin, Stern, and Rubin}{Gelman
  et~al.}{1995}]{gelman95bda}
Gelman, A., J.~B. Carlin, H.~S. Stern, and D.~B. Rubin (1995).
\newblock {\em Bayesian Data Analysis\/} (1st ed.).
\newblock CRC Press, London.

\bibitem[\protect\citeauthoryear{Gelman, Jakulin, Pittau, and Su}{Gelman
  et~al.}{2008}]{gelman2008weakly}
Gelman, A., A.~Jakulin, M.~G. Pittau, and Y.-S. Su (2008).
\newblock A weakly informative default prior distribution for logistic and
  other regression models.
\newblock {\em The Annals of Applied Statistics\/}~{\em 2\/}(4), 1360--1383.

\bibitem[\protect\citeauthoryear{Gelman and Little}{Gelman and
  Little}{1997}]{gelman:little-97}
Gelman, A. and T.~C. Little (1997).
\newblock Poststratification into many categories using hierarchical logistic
  regression.
\newblock {\em Survey Methodology\/}~{\em 23}, 127--135.

\bibitem[\protect\citeauthoryear{Ghitza and Gelman}{Ghitza and
  Gelman}{2013}]{Ghitza:gelman-13}
Ghitza, Y. and A.~Gelman (2013).
\newblock Deep interactions with {MRP}: {E}lection turnout and voting patterns
  among small electoral subgroups.
\newblock {\em American Journal of Political Science\/}~{\em 57\/}(3),
  762--776.

\bibitem[\protect\citeauthoryear{Good}{Good}{1967}]{Good1967}
Good, I.~J. (1967).
\newblock A {B}ayesian significance test for multinomial distributions.
\newblock {\em Journal of the Royal Statistical Society Series B\/}~{\em
  29\/}(3), 399--431.

\bibitem[\protect\citeauthoryear{Goodman}{Goodman}{1968}]{goodman1968analysis}
Goodman, L.~A. (1968).
\newblock The analysis of cross-classified data: Independence,
  quasi-independence, and interactions in contingency tables with or without
  missing entries: Ra fisher memorial lecture.
\newblock {\em Journal of the American Statistical Association\/}~{\em
  63\/}(324), 1091--1131.

\bibitem[\protect\citeauthoryear{Greenland}{Greenland}{2007}]{Greenland07}
Greenland, S. (2007).
\newblock Prior data for non-normal priors.
\newblock {\em Statistics in Medicine\/}~{\em 26}, 3578--3590.

\bibitem[\protect\citeauthoryear{Heeringa, Berglund, West, Mellipil{\'a}n, and
  Portier}{Heeringa et~al.}{2015}]{Heeringa15}
Heeringa, S.~G., P.~A. Berglund, B.~T. West, E.~R. Mellipil{\'a}n, and
  K.~Portier (2015).
\newblock Attributable fraction estimation from complex sample survey data.
\newblock {\em Annals of Epidemiology\/}~{\em 25\/}(3), 174--178.

\bibitem[\protect\citeauthoryear{Horvitz and Thompson}{Horvitz and
  Thompson}{1952}]{ht52}
Horvitz, D.~G. and D.~J. Thompson (1952).
\newblock A generalization of sampling without replacement from a finite
  university.
\newblock {\em Journal of the American Statistical Association\/}~{\em
  47\/}(260), 663--685.

\bibitem[\protect\citeauthoryear{Ireland and Kullback}{Ireland and
  Kullback}{1968}]{ireland1968contingency}
Ireland, C.~T. and S.~Kullback (1968).
\newblock Contingency tables with given marginals.
\newblock {\em Biometrika\/}~{\em 55\/}(1), 179--188.

\bibitem[\protect\citeauthoryear{Knuiman and Speed}{Knuiman and
  Speed}{1988}]{Speed88}
Knuiman, M.~W. and T.~P. Speed (1988).
\newblock Incorporating prior information into the analysis of contingency
  tables.
\newblock {\em Biometrics\/}~{\em 44\/}(4), 1061--1071.

\bibitem[\protect\citeauthoryear{Kott}{Kott}{2009}]{calibration:kott09}
Kott, P. (2009).
\newblock Calibration weighting: combining probability samples and linear
  prediction models.
\newblock In D.~Pfeffermann and C.~R. Rao (Eds.), {\em Handbook of Statistics,
  Sample Surveys: Design, Methods and Application}, Volume 29B. Elsevier.

\bibitem[\protect\citeauthoryear{Kott}{Kott}{2006}]{Kott-SM06}
Kott, P.~S. (2006).
\newblock Using calibration weighting to adjust for nonresponse and coverage
  errors.
\newblock {\em Survey Methodology\/}~{\em 32\/}(2), 133--142.

\bibitem[\protect\citeauthoryear{Kunihama and Dunson}{Kunihama and
  Dunson}{2013}]{Kunihama13}
Kunihama, T. and D.~B. Dunson (2013).
\newblock Bayesian modeling of temporal depen- dence in large sparse
  contingency tables.
\newblock {\em Journal of the American Statistical Association\/}~{\em
  108\/}(504), 1324--1338.

\bibitem[\protect\citeauthoryear{Laird}{Laird}{1978}]{Laird78}
Laird, N.~M. (1978).
\newblock Empirical {B}ayes methods for two-way contingency tables.
\newblock {\em Biometrika\/}~{\em 65}, 581--590.

\bibitem[\protect\citeauthoryear{Lazar, Meeden, and Nelson}{Lazar
  et~al.}{2008}]{lazar2008}
Lazar, R., G.~Meeden, and D.~Nelson (2008).
\newblock A noninformative {B}ayesian approach to finite population sampling
  using auxiliary variables.
\newblock {\em Survey Methodology\/}~{\em 34\/}(1), 51--64.

\bibitem[\protect\citeauthoryear{Little}{Little}{1983}]{little83-pi}
Little, R. (1983).
\newblock Comment on ``{A}n evaluation of model-dependent and
  probability-sampling inferences in sample surveys", by {M. H. H}ansen, {W. G.
  M}adow and {B. J. T}epping.
\newblock {\em Journal of the American Statistical Association\/}~{\em 78},
  797--799.

\bibitem[\protect\citeauthoryear{Little}{Little}{1991}]{little91}
Little, R. (1991).
\newblock Inference with survey weights.
\newblock {\em Journal of Official Statistics\/}~{\em 7}, 405--424.

\bibitem[\protect\citeauthoryear{Little}{Little}{1993}]{little93}
Little, R. (1993).
\newblock Post-stratification: A modeler's perspective.
\newblock {\em Journal of the American Statistical Association\/}~{\em 88},
  1001--1012.

\bibitem[\protect\citeauthoryear{Little}{Little}{2011}]{CalibratedBayes:Little11}
Little, R. (2011).
\newblock Calibrated {B}ayes, for statistics in general, and missing data in
  particular.
\newblock {\em Statistical Science\/}~{\em 26}, 162--174.

\bibitem[\protect\citeauthoryear{Little and Rubin}{Little and
  Rubin}{2002}]{Little-MDBook02}
Little, R. and D.~B. Rubin (2002).
\newblock {\em Statistical Analysis with Missing Data}.
\newblock Wiley.

\bibitem[\protect\citeauthoryear{Little and Wu}{Little and
  Wu}{1995}]{littlewu91}
Little, R. and M.-M. Wu (1995).
\newblock Models for contingency tables with known margins when target and
  samples populations differ.
\newblock {\em Journal of the American Statistical Association\/}~{\em
  86\/}(413), 87--95.

\bibitem[\protect\citeauthoryear{Lumley}{Lumley}{2016}]{lumley2016package}
Lumley, T. (2016).
\newblock survey: analysis of complex survey samples.
\newblock R package version 3.31-5.

\bibitem[\protect\citeauthoryear{Meng and Rubin}{Meng and
  Rubin}{1993}]{meng:rubin93}
Meng, X.-L. and D.~B. Rubin (1993).
\newblock Maximum likelihood estimation via the {ECM} algorithm: {A} general
  framework.
\newblock {\em Biometrika\/}~{\em 80\/}(2), 267--278.

\bibitem[\protect\citeauthoryear{Park, Gelman, and Bafumi}{Park
  et~al.}{2005}]{park:gelman:bafumi-04}
Park, D.~K., A.~Gelman, and J.~Bafumi (2005).
\newblock State-level opinions from national surveys: {P}oststratification
  using multilevel logistic regression.
\newblock In J.~E. Cohen (Ed.), {\em Public Opinion in State Politics}.
  Standord University Press.

\bibitem[\protect\citeauthoryear{Rao and Molina}{Rao and Molina}{2015}]{rao15}
Rao, J. and I.~Molina (2015).
\newblock {\em Small Area Estimation}.
\newblock John Wiley \& Sons, Inc.

\bibitem[\protect\citeauthoryear{{RTI International}}{{RTI
  International}}{2019}]{sudaan}
{RTI International} (2019, Apr).
\newblock Sudaan 11.0.3.
\newblock http://sudaansupport.rti.org/sudaan/index.cfm.

\bibitem[\protect\citeauthoryear{Rubin}{Rubin}{1983}]{rubin83-pi}
Rubin, D.~B. (1983).
\newblock Comment on ``{A}n evaluation of model-dependent and
  probability-sampling inferences in sample surveys," by {M. H. H}ansen, {W. G.
  M}adow and {B. J. T}epping.
\newblock {\em Journal of the American Statistical Association\/}~{\em 78},
  803--805.

\bibitem[\protect\citeauthoryear{S{\"a}rndal, Swensson, and
  Wretman}{S{\"a}rndal et~al.}{1992}]{modelass-sarndal92}
S{\"a}rndal, C.-E., B.~Swensson, and J.~H. Wretman (1992).
\newblock {\em Model Assisted Survey Sampling}.
\newblock Springer, New York.

\bibitem[\protect\citeauthoryear{Schafer}{Schafer}{1997}]{schafer:97}
Schafer, J.~L. (1997).
\newblock {\em Analysis of Incomplete Multivariate Data}.
\newblock London: Chapman \& Hall.

\bibitem[\protect\citeauthoryear{Schifeling and Reiter}{Schifeling and
  Reiter}{2016}]{reiter:ba16}
Schifeling, T.~A. and J.~P. Reiter (2016).
\newblock Incorporating marginal prior information in latent class models.
\newblock {\em Bayesian Analysis\/}~{\em 11\/}(2), 499--518.

\bibitem[\protect\citeauthoryear{Schouten}{Schouten}{2018}]{Schouten18}
Schouten, B. (2018).
\newblock Statistical inference based on randomly generated auxiliary
  variables.
\newblock {\em Journal of the Royal Statistical Society, Statistical
  Methodology, Series B\/}~{\em 80\/}(1), 33--56.

\bibitem[\protect\citeauthoryear{Si}{Si}{2019}]{rake:code19}
Si, Y. (2019).
\newblock Codes for manuscript "bayes-raking: Bayesian finite population
  inference with known margins".
\newblock https://github.com/yajuansi-sophie/raking.

\bibitem[\protect\citeauthoryear{Si and Gelman}{Si and
  Gelman}{2014}]{RHweighting}
Si, Y. and A.~Gelman (2014).
\newblock Survey weighting for {New York Longitudinal Survey on Poverty
  Measure}.
\newblock Technical report, Columbia University.

\bibitem[\protect\citeauthoryear{Si, Pillai, and Gelman}{Si
  et~al.}{2015}]{BNFP:SI15}
Si, Y., N.~S. Pillai, and A.~Gelman (2015).
\newblock Bayesian nonparametric weighted sampling inference.
\newblock {\em Bayesian Analysis\/}~{\em 10\/}(3), 605--625.

\bibitem[\protect\citeauthoryear{Si, Trangucci, Gabry, and Gelman}{Si
  et~al.}{2018}]{prior-si2018}
Si, Y., R.~Trangucci, J.~S. Gabry, and A.~Gelman (2018).
\newblock Bayesian hierarchical weighting adjustment and survey inference.
\newblock Survey Methodology (revision invited);
  https://arxiv.org/abs/1707.08220.

\bibitem[\protect\citeauthoryear{{Stan Development Team}}{{Stan Development
  Team}}{2017}]{stan-manual:2013}
{Stan Development Team} (2017).
\newblock Stan modeling language user's guide and reference manual.
\newblock http://mc-stan.org.

\bibitem[\protect\citeauthoryear{{Stan Development Team}}{{Stan Development
  Team}}{2018}]{stan-software:2013}
{Stan Development Team} (2018).
\newblock Stan: {A} {C}++ library for probability and sampling.
\newblock http://mc-stan.org.

\bibitem[\protect\citeauthoryear{Stephan}{Stephan}{1942}]{itf:stephan42}
Stephan, F.~F. (1942).
\newblock An iterative method of adjusting sample frequency tables when
  expected marginal totals are known.
\newblock {\em The Annals of Mathematical Statistics\/}~{\em 13\/}(2),
  166--178.

\bibitem[\protect\citeauthoryear{Vehtari, Gelman, and Gabry}{Vehtari
  et~al.}{2017}]{loo17}
Vehtari, A., A.~Gelman, and J.~S. Gabry (2017).
\newblock Practical bayesian model evaluation using leave-one-out
  cross-validation and {WAIC}.
\newblock {\em Statistics and Computing\/}~{\em 27}, 1413--1432.

\bibitem[\protect\citeauthoryear{Wimer, Garfinkel, Gelblum, Lasala, Phillips,
  Si, Teitler, and Waldfogel}{Wimer et~al.}{2014}]{RHreport}
Wimer, C., I.~Garfinkel, M.~Gelblum, N.~Lasala, S.~Phillips, Y.~Si, J.~Teitler,
  and J.~Waldfogel (2014).
\newblock Poverty tracker---monitoring poverty and well-being in {NYC}.
\newblock Columbia Population Research Center and Robin Hood Foundation.

\bibitem[\protect\citeauthoryear{Zheng and Little}{Zheng and
  Little}{2005}]{zhenglittle05}
Zheng, H. and R.~Little (2005).
\newblock Inference for the population total from
  probability-proportional-to-size samples based on predictions from a
  penalized spline nonparametric model.
\newblock {\em Journal of Official Statistics\/}~{\em 21}, 1--20.

\bibitem[\protect\citeauthoryear{Zheng and Little}{Zheng and
  Little}{2003}]{zhenglittle03}
Zheng, H. and R.~J. Little (2003).
\newblock Penalized spline model-based estimation of the finite populations
  total from probability-proportional-to-size samples.
\newblock {\em Journal of Official Statistics\/}~{\em 19\/}(2), 99--107.

\end{thebibliography}
\bibliographystyle{chicago}

\end{document}